\documentclass[twoside,11 pt]{article}

\makeatletter
\newcommand*{\rom}[1]{\expandafter\@slowromancap\romannumeral #1@}
\makeatother
\usepackage{amsmath}
\usepackage{mathtools}
\usepackage{amssymb}
\usepackage{mathrsfs}

\usepackage{graphicx}

\usepackage{cite}

\usepackage{color} 

\usepackage[labelfont=bf,labelsep=period,justification=raggedright]{caption}

\makeatletter
\renewcommand{\@biblabel}[1]{\quad#1.}
\makeatother

\date{}

\begin{document}

\begin{flushleft}
{\Large
\textbf{Generalization of neuron network model with delay feedback}
}
\bigskip
\\
Sanjeet Maisnam$^{a,b}$, R.K. Brojen Singh$^{b\ast}$
\\
\bigskip
$^a$Indian Institute of Science Education and Research, Kolkata-741246, India.\\
$^b$School of Computational and Integrative Sciences, Jawaharlal Nehru University, New Delhi-110067, India.\\
\bigskip
$^\ast$Corresponding Author
\bigskip
\end{flushleft}

\section*{Abstract}
We present generalized delayed neural network (DNN) model with positive delay feedback and neuron history. The local stability analysis around trivial local equilibria of delayed neural networks has applied and determine the conditions for the existence of zero root. We develop few innovative delayed neural network models in different dimensions through transformation and extension of some existing models. We found that zero root can have multiplicty two under certain conditions. We further show how the characteristic equation can have zero root and its multiplicity is dependent on the conditions undertaken. Finally, we generalize the neural network of $N$ neurons through which we determine the general form of Jacobian of the linear form and corresponding characteristic equation of the system. 

\bigskip

\section*{Introduction}
\label{abc}

The stability analysis in various dynamical systems allows us to understand near equilibrium properties of flows of the systems under pertubations and stability criteria around the equilibria. An equilbrium is locally stable if a system near the equilibrium approaches towards it \cite{Ott}, and will exhibit critical behavior of the system near that equilibrium point but may be different for different local equilibrium points. The local stability analysis, in general refers to the procedure of approximating a non-linear model by a linear one in the vicinity of local equilibrium point and analyzing the behavior of the linear form of the system. The local stability of the system can be characterized and analyzed by determining the eigenvalues of the linear system calculated from the so-called Jacobian matrix of the system \cite{Ott}.

The stability analysis have been applied to various dynamical systems such as exponential and logistic models of population growth, models of natural selection, delayed models of neural netwroks \cite{Ott} etc. The dynamics of natural systems (physical, chemical, biological systems) involves with delay time in the process, and can be modeled by delayed differential equations (DDEs) which allows to understand the important role of delay time, and have been an extensive area of research since the past few decades in a wide range of applications. The most fundamental functional differential equation (FDE) is the linear first order DDE which is of the form \cite{Fal}
\begin{eqnarray}
\label{109}
\frac{d{\bf x}}{dt} = f_1(t){\bf x}(t) + f_2(t){\bf x}(t - \tau) , \hspace{3mm} t > 0 
\end{eqnarray}
where, $f_1$ and $f_2$ are time dependent functions. $\tau$ is the time delay involved in the time evolution of vector ${\bf x}$. Such mathematical modeling is very useful in describing neuron interactions with different time delays which are generally reflected in experimental data such as EEG, EMG data and fMRI, DTI data etc. In such situation, the time delays involved in DDEs have been incorporated in the mathematical models of delayed neural networks which represent the propagation time among the axons and the dendrites. For coupled system of DNNs, the dynamics involves the excitatory and inhibitory influences and the activation level of one neuron is affected by the local feedback from the other neurons \cite{Xin}.
However, to gain a better insight into such non-equilibrium dynamical systems, one needs to focus on the equilibria and its stability. 

The delayed systems of differential equations modeling of neural networks have been a cutting-edge research in the area of neuroscience. Both single variable and multivariate discrete and continuous time models for DNNs have been developed and investigated in different dimensions. For instance, Liao et al. \cite{Xia} studied a single delayed neuron model, while elaborate discussions of two-neuron model was discussed in different works \cite{Gop,Che,Gia,Sha,Wei}. Further, tri-neurons networks have also been studied with different feedback models \cite{Xin,Fal,Gup,Guo,JWe,and,Son} and four-neuron BAM \cite{Juh,Wen,Wan} models are discussed upto some extent. However, there has not been a generalised model of neural network from which a simpler model can be deduced, and the general Jacobian and the corresponding characteristic equation are still left as open questions. We focus on generalization of the models to N-dimensions to present the outsets of these innovative models and demonstrate the corresponding generalisations of the underlying properties. The perspective for the extension or generalisation is, however, referred to some of the existing models.
 
In this paper, we focus on the linear stability analysis of DNNs by linearising the model equations around the trivial local equilibria. Initially, the typical non-linear function `tanh' is associated with the DDEs of the models which is then expanded using the Taylor series expansion and the higher order terms are truncated during the linearisation leaving only the linear terms. We consider the Jacobian, \Big|$\frac{\partial(x_1,..., x_n)}{\partial(C_1,..., C_n)}$\Big| = 0, where $n$ is the number of neurons, to obtain the characteristic equation of the system and determine its zero root. This brief delineates the conditions for the subsistence of multiplicity 2 of the zero root, as this leads to the Bogdanov-Takens bifurcation \cite{Gui, Xin}. One can use the centre manifold reduction \cite{Far,Mag,Jia,Cam,Gil,Qes} and the normal form method \cite{Ria,Cho,Guc,Kuz} to compute the simpler normal form of the original DDEs and analyse the dynamic behaviors of the system DNNs \cite{Xin}. Furthermore, we observe similarity in the Jacobians and the corresponding characteristic equations of the systems and from this standpoint, we generalize the form of the Jacobian and the corresponding characteristic equation to $N$ dimensions. 

\section*{Local Stability Analysis}
\label{def}

The local or linear stability analysis of DDE can be done by taking into account the linear form of the non-linear DDEs around any local equilibrium (\ref{109}). This linearisation is achieved by Taylor expansion of the non-linear function involved in DDEs. To deal with such equations for the non-trivial solutions, one assumes that the solution exists in the linear form $x$ = $C e^{\lambda t}$ , $C$ $\in$ $\mathbb{C}$(complex space) and the model DDEs will reduce to characteristic equation of the system (\ref{109}) \cite{Fal, Sun}. We consider the DDEs which are constant-coefficient equations of the following form,
\begin{eqnarray}
\label{110}
\frac{dx}{dt} = f_1x(t) + f_2x(t - \tau) , \hspace{3mm} t > 0 
\end{eqnarray}
\noindent The following the linearization process of the above form of equations, one can find the the following characteristic equation:
\begin{eqnarray}
\label{108}
(\lambda - f_1) - f_2 e^{- \lambda \tau} = 0
\end{eqnarray}
Analyzing the above equation (\ref{108}), we check for the existence of the zero root and the necessary conditions for the subsistence of its multiplicity. 

\subsubsection*{1D Dynamic Neural Network model}
\label{jkl}

The one dimensional neural network (NN) which we consider here is slightly different from that investigated by Liao et al. in \cite{Xia} in the arguments of the non-linear function in the model. The model is the simplication of the model considered in \cite{Gui} which is discussed briefly in the above section. The model equation is given by
\begin{eqnarray}
\label{200}
\frac{dx}{dt} = - x(t) + \alpha \hspace{1mm} \textrm{tanh}(x(t - \tau_f))
\end{eqnarray}
\noindent  where $\alpha$ and $\tau_f$ have the usual meanings. We study the linear stability analysis in the vicinity of the trivial equilibrium of the system (\ref{200}), assuming that it always possesses an equilibrium at the origin and the solution exists in the linear form of $x$ = $Ce^{\lambda t}$, where $\lambda$ is the eigenvalue and $C$ $\in$ $\mathbb{C}$. Using Taylor series approximation on the sigmoid function 'tanh' around the trivial equilibrium reduces to the linear form $\frac{dx}{dt} = - x(t) + \alpha x(t - \tau_f)$. The corresponding characteristic equation can be obtained as,
\begin{eqnarray}
\label{115}
F^{(0)}(\lambda)=(\lambda + 1) - \alpha e^{- \lambda \tau_f} = 0
\end{eqnarray}
where, $F^{(0)}$ is the functional form of order 0.

\vspace{4mm}
\noindent \textbf{Lemma 1.1}: If $x_0$ is a root of $F^{(0)}(x)$ of multiplicity $r$, then $F^{(1)}(x)$ also has $x_0$ as its root of multiplicity $r-1$ \cite{Dic}. In other words, a root of $F^{(0)}(x)$ having a multiplicity $r$ is a single root of $F^{(r)}(x)$ and not a root of $F^{(r+1)}(x)$.

\vspace{4mm}
\noindent \textbf{Lemma 1.2}: $\lambda$ = 0 is a single root of (\ref{115}) $\emph{iff}$ $\alpha$ = 1. 

\vspace{3mm}
\noindent $\emph{Proof}$ \hspace{3mm}  Suppose $\lambda$ = 0 is a single root of (\ref{115}). Then,
\begin{eqnarray}
F(0) = 1 - a = 0
\end{eqnarray}
\noindent which gives a = 1. For the other direction, by Lemma 1.1, we need to claim $F^{(0)}(0) = 0$ and $F^{(1)}(0) \neq 0$ given that $\alpha$ = 1. Clearly, $F^{(0)}(0) = 0$ and $F^{(1)}(0) = - \tau_f$.

\subsubsection*{2D Dynamic Neural Network model}
\label{mno}

We consider the Delayed NN model considered by Fan et al. in \cite{Gui} which is different from the dynamic behaviors of two-neuron networks studied in \cite{Gop,Che,Gia,Sha,Wei}. The network is being modeled, along with the introduction of delayed self-feedback and a delayed connection from the other neuron, by a coupled system of DDEs as follows:
\begin{eqnarray}
\label{111}
\frac{dx_1}{dt}&=& - x_1(t) + \alpha \hspace{1mm} \textrm{tanh}(x_1(t - \tau_f)) - \alpha_{12} \textrm{tanh}(x_2(t - \tau_2)), \nonumber \\
\frac{dx_2}{dt}&=& - x_2(t) - \alpha \hspace{1mm} \textrm{tanh}(x_2(t - \tau_f)) + \alpha_{21} \textrm{tanh}(x_1(t - \tau_1))
\end{eqnarray} 
\noindent where $a_{12}$ and $a_{21}$ are the connection strengths with $\tau_i$ ($i$ = 1, 2) as the connection delays and $\alpha$ $>$ 0 is the feedback strength having a delay $\tau_f$. The linearized form of (\ref{111}) is given by,
\begin{eqnarray}
\label{112}
\frac{dx_1}{dt}&=&-x_1(t) + \alpha x_1(t - \tau_f) - \alpha_{12} x_2(t - \tau_2), \nonumber \\
\frac{dx_2}{dt}&=&- x_2(t) - \alpha x_2(t - \tau_f) + \alpha_{21} x_1(t - \tau_1)
\end{eqnarray} 
The corresponding characteristic equation calculated from the Jacobian of equation (\ref{112}) in $\lambda$ is given by,
\begin{eqnarray}
\label{113}
 F(\lambda) = (1 + \lambda)^2 - \alpha^2 e^{-2 \tau_f \lambda} + \alpha_{12}\alpha_{21} e^{2 \tau \lambda} = 0
\end{eqnarray}
where, $\tau = \frac{\tau_1 + \tau_2}{2}$. 
\vskip 0.7cm

\noindent \textbf{Theorem 2.1}: \hspace{3mm}($i$) The characteristic equation (\ref{113}) has a zero root ($\lambda$ = 0) of multiplicity 2 \emph{iff} \cite{Gui}
\begin{eqnarray}
\label{114}
\tau > \tau_f, \hspace{3mm} \textrm{and} \hspace{3mm} \alpha^2 = \frac{\tau + 1}{\tau - \tau_f} , \hspace{3mm} \alpha_{12}\alpha_{21} = \frac{\tau_f + 1}{\tau - \tau_f}
\end{eqnarray}
($ii$) The maximum multiplicity of the zero root is 2.

\vspace{3mm}
$\emph{Proof}$ \hspace{3mm}($i$) First suppose that $\lambda$ = 0 is a root of (\ref{113}) of multiplicity 2. This results into the equations:
\begin{eqnarray}
F(0) = 0 \hspace{5mm} \textrm{and} \hspace{5mm} F^{(1)}(0) =0 \nonumber
\end{eqnarray}
\begin{eqnarray}
\Rightarrow 1 - \alpha^2 + \alpha_{12}\alpha_{21} = 0 \hspace{5mm} \textrm{and} \hspace{5mm} 1 + \alpha^2 \tau_f - \alpha_{12}\alpha_{21} \tau = 0 \nonumber
\end{eqnarray}
\noindent Solving these equations yields the conditions (\ref{114}).

For the other direction, by Lemma \ref{def}.1, it would be suffice to show $F^{(0)}(0)$ = 0, $F^{(1)}(0)$ =0 and $F^{(2)}(0)$ $\neq$ 0 successively when $\alpha^2$ and $\alpha_{12}\alpha_{21}$ satisfy the conditions (\ref{114}). Eventually, $F^{(0)}(0)$ = 0, $F^{(1)}(0)$ =0 and $F^{(2)}(0)$ = 1 + 2($\tau$ + $\tau_f$).

($ii$) We take into account the method of contradiction to prove this statement. Suppose $\lambda$ = 0 is a multiple root of (\ref{113}) having multiplicity 3. This necessarily implies $F^n(0)$ = 0 ($n$ = 0, 1, 2), where $n$ denotes the number of derivatives w.r.t. $\lambda$. Substituting the values of $\alpha^2$ and $\alpha_{12}\alpha_{21}$ in $F^{(2)}(0)$ = 0, we obtain $\tau$ + $\tau_f$ = - $\frac{1}{2}$. However, $\tau$ $>$ $\tau_f$ and $\tau$ $>$ 0. This contradicts our supposition and hence the proof is complete.

\subsubsection*{3D Dynamic Neural Network model}
\label{pqr}

In three neurons model, each neuron has the ability to activate itself and the activation is dependent on the history of its previous activation \cite{Xin}. The axonal and dendritic propagation time, also called the synaptic delay, is considered associated with the local positive feedback, biologically termed `reverberation' \cite{Lia}. The model is described by,
\begin{eqnarray}
\label{116}
\frac{dx_i}{dt} = - x_i(t) + \alpha_i \textrm{tanh}[x_{i+2} - \beta x_{i+2}(t - \tau)], \hspace{3mm} \tau > 0
\end{eqnarray}
\noindent where $i=1,2,3$, $x_i$ represents the activation level of the $i^{th}$ neuron with the activity coefficient $\alpha_i$, $\beta$ denotes the inhibitory influence measure of the past history, $x_{i+2}$ is the reverberation. The linearization of equation (\ref{116}) allows to obtain the following equation, 
\begin{eqnarray}
\label{117}
\dot{x}_i(t) = - x_i(t) + \alpha_i [x_{i+2} - \beta x_{i+2}(t - \tau)]
\end{eqnarray}
and the corresponding characteristic equation at the equilibrium point (origin) can be obtained as below,
\begin{eqnarray}
\begin{vmatrix}  \lambda + 1 & 0 & - \alpha_1 (1 - \beta e^{- \lambda \tau}) \\ 
- \alpha_2 (1 - \beta e^{- \lambda \tau}) & \lambda + 1 & 0 \\
0 & - \alpha_3 (1 - \beta e^{- \lambda \tau}) & \lambda +1 \end{vmatrix} = 0 \nonumber
\end{eqnarray}
\noindent which reduces to the following functional form,
\begin{eqnarray}
\label{118} 
F(\lambda) = {(\lambda + 1)}^3 - \alpha_{123} {(1 - \beta e^{- \lambda \tau})}^3 = 0 ; \hspace{3mm} \alpha_{ijk...} = \alpha_i \alpha_j \alpha_k ....
\end{eqnarray}
\noindent Thus, the characteristic equation (\ref{118}) contains the terms $(\lambda + 1)$ and $(1 - \beta e^{- \lambda \tau})$ with degree 3. If $\beta$ = 1 - $\frac{1}{\alpha}$ and 1 $<$ $\alpha$ $<$ 4, where $\alpha$ = $\sqrt[3]{\alpha_{123}}$, then (\ref{118}) has a zero root of multiplicity two and no purely imaginary roots. The conditions of single root and double root are briefly discussed in \cite{Xin}.

\subsection*{4D Dynamic Neural Network model}
\label{stu}
\begin{figure}
\hspace{28mm}\includegraphics[width=10cm,height = 8cm,angle = 0]{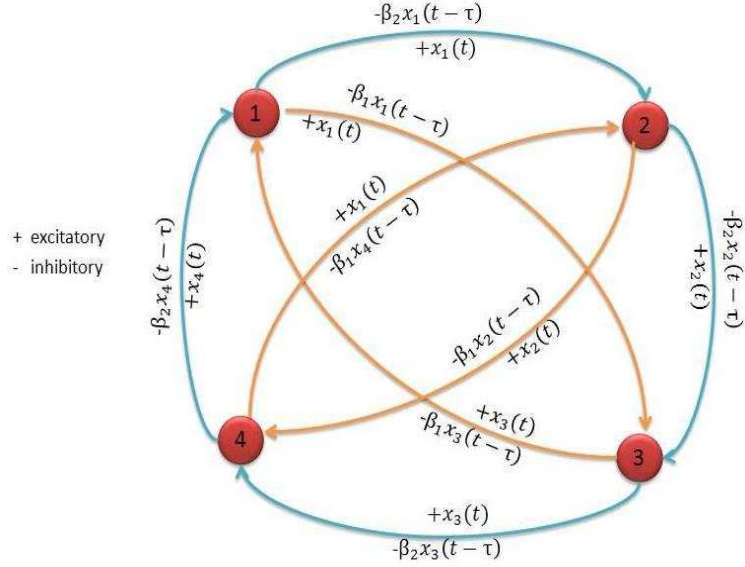}
\caption{Architecture of the four-neuron model}
\label{plot}
\end{figure}

We then present an innovative approach to the delayed networks containing four neurons, by extending the 3D DNN model. The features and the forms of equations are equivalent, except for the introduction of cross-linking among the neurons. As a result, there arises different measures of the inhibitory influence of the past history as shown in Fig. (\ref{plot}). The delayed bidirectional associative memory(BAM) neural network of four neurons with time delays have been investigated extensively in \cite{Juh,Wen,Wan}.

\noindent The different measures signifies that the inhibitory influences vary with different time delays. Moreover, the local positive feedback is also increased. The model is given by,
\begin{eqnarray}
\label{119}
\frac{dx_i}{dt} = - x_i(t) + \alpha_i \textrm{tanh} \Big[x_{i+2}(t) - \beta_1 x_{i+2}(t- \tau_1)\Big] + \alpha_i \textrm{tanh}\Big[x_{i+3}(t) - \beta_2 x_{i+3}(t - \tau_2) \Big]
\end{eqnarray}
where $i=1,2,3,4$, $\beta_1$ and $\beta_2$ are the measures of inhibitory influences during time delays $\tau_1$ and $\tau_2$ respectively, $x_{i +2}$ and $x_{i+3}$ are the feedback, and $x_i$ and $\alpha_i$ have the usual meanings. The linearization of (\ref{119}) and keeping $\tau_1$ = $\tau_2$ = $\tau$, we get the following linearized form,
\begin{eqnarray}
\label{120}
\frac{dx_i}{dt}= - x_i(t) + \alpha_i \Big[x_{i+2}(t) - \beta_1 x_{i+2}(t- \tau)\Big] + \alpha_i \Big[x_{i+3}(t) - \beta_2 x_{i+3}(t - \tau) \Big] +O(t^2)
\end{eqnarray}
The corresponding Jacobian matrix of the model is given by,
\begin{eqnarray}
\begin{vmatrix}  \lambda + 1 & 0 & - \alpha_1 f_1 & - \alpha_1 f_2 \\ 
- \alpha_2 f_2 & \lambda + 1 & 0 & - \alpha_2 f_1 \\
- \alpha_3 f_1 & - \alpha_3 f_2 & \lambda + 1 & 0 \\
0 & - \alpha_4 f_1 & - \alpha_4 f_2 & \lambda +1 \end{vmatrix} = 0 \nonumber
\end{eqnarray}
where, $f_j(\tau)$ = $(1 - \beta_j e^{-\lambda \tau})$. Expanding the determinant, we arrive at the following characteristic equation,
\begin{eqnarray}
F(\lambda) = (\lambda + 1)^4 - \alpha_{1234} f_2^4 + g(\lambda, \tau) \nonumber 
\end{eqnarray} 
where, the function $g(\lambda,\tau)$ is given by,
\begin{eqnarray}
g(\lambda,\tau)&=&(\lambda + 1)^2 f_1^2 (\alpha_{13}-\alpha_{24})-(\lambda + 1)f_1f_2[\alpha_{124} + f_2(\alpha_{123} + \alpha_{134})] \nonumber \\
&&+ [\alpha_{1234}f_2^3 - \alpha_{234}f_1f_2^2]  
\end{eqnarray}

\vspace{4mm}
\noindent \textbf{Theorem 4.1}: Suppose $\beta_1$ = 1 and $\beta_2$ = 1 + ${\epsilon_4}^{-\frac{1}{3}}$ , then \\

($i$) $\lambda$ = 0 is a zero root of multiplicity 1 \emph{iff} $\tau$ $\neq$ $\frac{4}{3 {\epsilon_4}^{\frac{1}{3}} + 3 + \epsilon_2 {\epsilon_4}^{-\frac{1}{3}} - {\epsilon_4}^{-\frac{2}{3}} (\epsilon_3 + \epsilon_5)}$ and \\

($ii$) $\lambda$ = 0 is a zero root of multiplicity 2 \emph{iff} $\tau$ = $\frac{4}{3 {\epsilon_4}^{\frac{1}{3}} + 3 + \epsilon_2 {\epsilon_4}^{-\frac{1}{3}} - {\epsilon_4}^{-\frac{2}{3}} (\epsilon_3 + \epsilon_5)}$, \\

\noindent where {$\epsilon_{1}$} = {$\alpha_{13} - \alpha_{24}$}, {$\epsilon_2$} = {$\alpha_{124}$}, {$\epsilon_3$} = {$\alpha_{123} + \alpha_{134}$}, {$\epsilon_4$} = {$\alpha_{1234}$} and {$\epsilon_5$} = {$\alpha_{234}$}. 
\\
\vspace{3mm}

\noindent \emph{Proof} \hspace{3mm} We claim the above theorem using Lemma 1.1. Clearly, 
{\small
\begin{eqnarray}
F(0) = 1 + (1 - \beta_1)^2 \epsilon_1 - (1 - \beta_1)(1 - \beta_2)[\epsilon_2 + (1 - \beta_2)\epsilon_3] 
+ [\epsilon_4(1 - \beta_1)^4 + \epsilon_4(1 - \beta_2)^3 - \epsilon_5(1- \beta_1)(1 - \beta_2)^2] \nonumber 
\end{eqnarray}  }
\noindent Substituting the aforementioned values of $\beta_1$ and $\beta_2$ ultimately yields $F(0) = 0$. And, 
{\small
\begin{eqnarray}
F^{(1)}(\lambda) = 4(\lambda + 1)^3 + (\lambda + 1)^2(1 - \beta_1 e^{- \lambda \tau})(2 \beta_1 \epsilon_1 \tau) +(\lambda + 1)\Big[2 \epsilon_1(1 - \beta_1 e^{- \lambda \tau})^2 - (-2 \tau \beta_1 \beta_2 e^{- 2 \lambda \tau} + \tau(\beta_1 \nonumber
\end{eqnarray}
\begin{eqnarray}
+ \beta_2)e^{-\lambda \tau})(\epsilon_2 + (1 - \beta_2 e^{- \lambda \tau})\epsilon_3) - (\beta_1 \beta_2 e^{- 2 \lambda \tau} - (\beta_1 + \beta_2) e^{- \lambda \tau} + 1)(\beta_2 \tau \epsilon_3 e^{- \lambda \tau}) \Big]  -\Big[(\beta_1 \beta_2 e^{- 2 \lambda \tau} - (\beta_1 +   \nonumber
\end{eqnarray}
\begin{eqnarray}
\beta_2) e^{- \lambda \tau} + 1)(\epsilon_2 + (1 - \beta_2 e^{- \lambda \tau} \epsilon_3 + 2 \epsilon_5 \beta_2 \tau) - 4\epsilon_4 \beta_1 \tau(1 - \beta_2 e^{- \lambda \tau})^3 - (1 - \beta_2 e^{- \lambda \tau})^2(3 \epsilon_4 \beta_2 \tau + \epsilon_5 \beta_1 \tau e^{- \lambda \tau}) \Big] \nonumber 
\end{eqnarray} }
{\small
\begin{eqnarray}
\Rightarrow F^{(1)}(0) = 4 + \tau \Big[-\epsilon_2 \epsilon_4^{-\frac{2}{3}} +  {\epsilon_4}^{-\frac{2}{3}}(\epsilon_3 - 3 \epsilon_4 (1 + \epsilon_4^{-\frac{1}{3}}) +  \epsilon_5) \Big] > 0 \nonumber
\end{eqnarray} } 
$\noindent$ If $\tau=$ $\frac{4}{3 {\epsilon_4}^{\frac{1}{3}} + 3 + \epsilon_2 {\epsilon_4}^{-\frac{1}{3}} - {\epsilon_4}^{-\frac{2}{3}} (\epsilon_3 + \epsilon_5)}$, it ultimately follows that $\tau \Big[-\epsilon_2 \epsilon_4^{-\frac{2}{3}} +  {\epsilon_4}^{-\frac{2}{3}}(\epsilon_3 - 3 \epsilon_4 (1 + \epsilon_4^{-\frac{1}{3}}) +  \epsilon_5) \Big] = -4$ and hence, $F^{(1)}(0) = 0$, $F^{(2)}(0)$ $\neq$ 0. This completes the proofs.

\subsection*{5D Dynamic Neural Network model}
\label{vwx}

We now present the 5D DNN model by extending the 4D DNN model. The increase of one neuron, however, adds up another cyclic pathway of neurons in a loop. As a result, there are increased influences and feedback from other neurons and their past history. The model DDEs are given by,
\begin{eqnarray}
\label{122}
\frac{dx_i}{dt} &=& - x_i(t) + \alpha_i \textrm{tanh} \Big[x_{i+2}(t) - \beta_1 x_{i+2}(t- \tau_1)\Big] + \alpha_i \textrm{tanh}\Big[x_{i+3}(t) - \beta_2 x_{i+3}(t - \tau_2) \Big] \nonumber\\
&& +\alpha_i \textrm{tanh} \Big[x_{i+4}(t) - \beta_3 x_{i+4}(t- \tau_3)\Big]
\end{eqnarray}
\noindent where $i=1,2,3,4,5$, $\beta_j$ measures the inhibitory influences during time delays $\tau_j$ respectively where $j$ = \{1, 2, 3\}; $x_{i+4}$ are the feedback, and other notations have the usual meanings. The linear form of (\ref{122}) is given by, 
\begin{eqnarray}
\label{123}
\dot{x}_i(t)&=&- x_i(t) + \alpha_i \Big[x_{i+2}(t) - \beta_1 x_{i+2}(t- \tau_1)\Big] + \alpha_i \Big[x_{i+3}(t) - \beta_2 x_{i+3}(t - \tau_2) \Big] 
\nonumber\\
&& + \alpha_i \Big[x_{i+4}(t) - \beta_3 x_{i+4}(t- \tau_3)\Big]
\end{eqnarray}
\noindent Assuming the delays are identical and the solutions exist in the linear form as in the previous models, the corresponding Jacobian is given by,
\begin{eqnarray}
\begin{vmatrix}  \lambda + 1 & 0 & - \alpha_1 f_1 & - \alpha_1 f_2 & -\alpha_1 f_3 \\ 
 - \alpha_2 f_3 & \lambda + 1 & 0 & - \alpha_2 f_1 & - \alpha_2 f_2 \\
 - \alpha_3 f_2 & - \alpha_3 f_3 & \lambda + 1 & 0 & - \alpha_3 f_1 \\
 - \alpha_4 f_1 & - \alpha_4 f_2 & - \alpha_4 f_3 & \lambda +1 & 0 \\
 0 & - \alpha_5 f_1 & - \alpha_4 f_2 & - \alpha_5 f_3 & \lambda + 1 \end{vmatrix} = 0 \nonumber
\end{eqnarray}
\noindent and this generates the characteristic equation of the system (\ref{123}) as follows,
\begin{eqnarray}
F(\lambda) = (\lambda + 1)^5 - \alpha_{12345} f_3^5 + h(\lambda, \tau) \nonumber 
\end{eqnarray}
where, the function $h(\lambda, \tau)$ is given,
\begin{eqnarray}
h(\lambda, \tau)&=& (\lambda + 1) \Big[(\lambda + 1)\Big\{(- \alpha_{345} f_1 f_3^2 - \alpha_{35} f_1 f_2 (\lambda + 1) \Big\} - \alpha_2 f_1 \Big\{\alpha_{34} f_3^3 (\lambda + 1) - \alpha_{345} f_2^2 f_1 \nonumber\\
&&+\alpha_{345} f_1^2 f_3 + (\lambda + 1)^2 \alpha_4 f_2 + \alpha_2 f_2 \Big\{ - \alpha_{345} f_3^3 - \alpha_5 (\lambda +1)^2 -\alpha_{35} f_2 f_3 (\lambda + 1) \nonumber\\
&&- \alpha_{45} f_2 f_3 (\lambda + 1) \Big\}\Big]-\alpha_1 f_1 \Big[ -\alpha_2 f_3 \Big\{ -(\lambda + 1)^2 \alpha_3 f_3 - \alpha_{34} f_1 f_2 f_3 - \alpha_{35} f_1^2 (\lambda  + 1) \Big\}  \nonumber\\
&&- (\lambda + 1) \Big\{ -(\lambda + 1)^2 \alpha_3 f_2 - \alpha_{345} f_1^2 f_3\Big\}  
- \alpha_2 f_1\Big\{\alpha_{34} f_2^2(\lambda  + 1)^2 - \alpha_{345} f_1^3 - \alpha_{34} f_1 f_3 \Big\} \nonumber \\
&&+ \alpha_1f_2 \Big\{ - \alpha_{345} f_2^2 f_3 - \alpha_{35} f_1 f_2 (\lambda  + 1) + \alpha_{345} f_1 f_3^2 \Big\} \Big] \nonumber \\
&&+\alpha_1 f_2 \Big[-\alpha_2 f_3 \Big\{(\lambda  + 1)\alpha_{34} f_3^2 - \alpha_{345} f_1 f_2^2 + \alpha_{345} f_1^2 f_3  + \alpha_4 f_2(\lambda  + 1)^2 \Big\} \nonumber \\
&&- (\lambda  + 1) \Big\{ (\lambda + 1) \alpha_{34} f_2 f_3 -\alpha_{345} f_1^3 - \alpha_{34} f_1 f_3 (\lambda  + 1) \Big\} + \alpha_2 f_2 \Big\{ - \alpha_{345} f_2^3  \nonumber \\
&&+ \alpha_{45} f_1^2 (\lambda  + 1) + \alpha_{345} f_1 f_2 f_3 + \alpha_{345} f_1 f_2 f_3 \Big\} \Big] -\alpha_1 f_3 \Big[-\alpha_2 f_3 \Big\{- \alpha_{345} f_1^3 \nonumber \\
&&- \alpha_5 f_1(\lambda  + 1)^2 - \alpha_{35} f_2 f_3 (\lambda  + 1)  - \alpha_{45} f_2 f_3 (\lambda  + 1) \Big\} - (\lambda  + 1) \Big\{ - \alpha_{345} f_2^2 f_3 \nonumber \\
&&- \alpha_{35} f_2^2 (\lambda + 1)  - \alpha_{45} f_1 f_3 (\lambda  + 1) \Big\} + \alpha_2 f_1 \Big\{ - \alpha_{345} f_2^3  + \alpha_{45} f_1^2 (\lambda  + 1) \nonumber \\
&&+ \alpha_{345} f_1 f_2 f_3 + \alpha_{345} f_1 f_2 f_3 \Big\} \Big]
\end{eqnarray}

\noindent Thus, in this model also, the characteristic equation contains the terms $(\lambda + 1)^n$, $(1- \beta e^{- \lambda \tau})^n$ associated with the product of all $\alpha_i$'s, and a function of $\lambda$ and $\tau$. The inhibitory influence $\beta$ represents the one measure during $\tau_n$ delay in the loop containing $n$ neurons. 

\subsection*{Generalization of the DNN Model}
\label{xyz}

We now generalized the dynamic neural network model by extending the state space to $N$ dimensions. We introduce the network of $N$ neurons with cross-linking among them, and the local feedback responses are made in cyclic order of $j+2$ neurons, where $j = 1,2, 3,..., N-2$. The synaptic delay of the neurons differs from each other, nevertheless, we consider the time delays to be identical in the analysis($\tau_j$ = $\tau$). We assume that the pathways within a cyclic network of neurons has the same measure of the inhibitory influences of the past history, $\beta_j$ over the loop, however, the measures remain varying over each and every cyclic network of $j+2$ neurons. Thus, the $N$-neuron model is described by,
\begin{eqnarray}
\label{121}
\dot{x}_i(t) = - x_i(t) + \alpha_i \displaystyle\sum_{j=1}^{N-2} \textrm{tanh} \Big[x_{i+j+1}(t) - \beta_j x_{i+j+1}(t - \tau)\Big]
\end{eqnarray}
\noindent where $i$(mod $N$) and the term symbols have their usual meanings. Similarly, using linearization process around the origin, DDE (\ref{121}) can be written as
\begin{eqnarray}
\label{125}
\dot{x}_i(t) = - x_i(t) + \alpha_i \displaystyle\sum_{j=1}^{n-2} \Big[x_{i+j+1}(t) - \beta_j x_{i+j+1}(t - \tau)\Big] 
\end{eqnarray}
The forms of the solutions are similar to those deployed in the previous sections. The Jacobian of (\ref{125}) has the following characteristics: \\
($i$) For lower triangular region of the matrix, each entry of the diagonal having ($n-j$) elements have the following form all the way alongside the main diagonal: 
\begin{eqnarray}
	\begin{cases}
\hspace{3mm} -\alpha_{i+1} (1 - \beta_{N-j-1} \hspace{1mm} e^{- \lambda \tau}) \hspace{2mm} \textrm{where} \hspace{2mm} j \leq i \leq N-1, \hspace{2mm} \textrm{for~each} \hspace{1mm} 1 \leq j \leq N-2  \\
\hspace{30mm} 0 \hspace{47.5mm}, \hspace{2.5mm} \textrm{for} \hspace{1mm} j = N-1 
	\end{cases}   \nonumber
\end{eqnarray}
\noindent ($ii$) For upper triangular region of the matrix, diagonal having ($n-j$) elements have the entries alongside the main diagonal in the form of
\begin{eqnarray}
	\begin{cases}
\hspace{3mm} -\alpha_{N-k} (1 - \beta_{j-1} \hspace{1mm} e^{- \lambda \tau}) \hspace{2mm} \textrm{where} \hspace{2mm} N-1 \leq k \leq j, \hspace{2mm} \textrm{for~each} \hspace{1mm} 2 \leq j \leq N-1  \\
\hspace{30mm} 0 \hspace{45mm}, \hspace{2mm} \textrm{for} \hspace{1mm} j = 1 
	\end{cases}   \nonumber
\end{eqnarray}
\noindent and \\
($iii$) For $j$ = 0, i.e. diagonal elements, each entry is of the form ($\lambda$ + 1) where $\lambda$ is the eigenvalue. 

\vspace{4mm}
\noindent \textbf{Theorem N.1}: The general characteristic equation for the linear form of the system of DNN exists in the form of 
\begin{eqnarray}
\label{124}
F(\lambda) = (\lambda + 1)^N + (-1)^N (1 - \beta e^{-\lambda \tau})^N \prod\limits_{i=1}^N \alpha_i + y(\lambda, \tau) 
\end{eqnarray}
where $\beta$ represents the measure of inhibitory influence during $\tau_N$ delay in the largest pathway containing all the $N$ neurons. The function $y(\lambda, \tau)$ has to be determined from the expansion of the Jacobian of N dimention.

\section{Discussions and Conclusions}
\label{zab}

Four new models of DNN, one each in 1-D, 4-D, 5-D and $N$-D, have been developed in this small piece of work. Our models(1-D, 4-D, 5-D and $n$-D) are the transformation and extensions of the existing models in \cite{Gui, Lia}. In each model, we perform the local stability analysis and consequently, we are able to determine the zero root of the characteristic equation of the linear form of each and every model equation, excluding the generalised model. The zero root has multiplicity 2 in 2-D, 3-D and 4-D under certain conditions. In 1-D model, the zero eigenvalue is a single root when the feedback strength is one unit. The generalised $n$-D model is an extension from the smaller 3-D, 4-D and 5-D models. From the analysis in these models, we can conclude that there always exist certain conditions, in every model of dimension $n$ $>$ 1, for the existence of the zero root and the subsistence of multiplicity 2. Moreover, we can observe a similar pattern in the characteristic equations of all the models, although one is different from each other owing to the forms of the equations. The general characteristic equation of DNN is very complicated to obtain as it is quite cumbersome to evaluate the value of higher order determinants analytically and there has not been a determinate form of the equation, although there are general formula like Leibnitz's rule, Laplace's formula, etc. As a result, we derive the general characteristic equation (\ref{124}) with some indeterminate terms expressed as function(s) of $\lambda$ and $\tau$ and show the similar terms involved. These findings may yield innovative results on the dynamics of neural networks and provide better insights into the mathematical models of DNNs.

\subsection*{Acknowledgement} We acknowledge for providing financial support from the CSIR, India, under sanction no. 25(0221)/13/EMR-II.

\end{document}